\documentclass[a4paper,10pt]{article}
\usepackage{
times,
amsmath,amssymb,epsfig,graphics,graphicx}

\newcommand{\be}{\begin{eqnarray}}
\newcommand{\ee}{\end{eqnarray}}

\begin{document}

\begin{center}
{\Large \bf  Second order differential realization of the Bargmann-Wigner framework for particles of any spin}
\end{center}

\vspace{0.51cm}

\begin{center}
V.\ M.\ Banda Guzm\'an$^*$ and M.\ Kirchbach\\
Institute of Physics, Autonomous University at San Luis Potosi,\\
Av. Manuel Nava 6, University Campus, San Luis Potosi, S.L.P. 78290, Mexico\\
e-mail:vmbg@ifisica.uaslp.mx
\end{center}

\vspace{0.51cm}
\begin{flushleft}
{\bf Abstract:} The Bargmann-Wigner (BW) framework describes particles of any spin-$j$ in terms of Dirac spinors of rank $2j$,
obtained as the local direct product of $n$ Dirac spinor copies, with $n=2j$.  Such spinors are reducible under Lorentz transformations, and contain among others also $(j,0)\oplus (0,j)$-pure spin representation spaces. The $2(2j+1)$ degrees of freedom of the latter are supposed to be identified by a projector, correspondingly  given by the $n$-fold direct product of the covariant parity projector within the Dirac spinor space.
Upon imposing on the spinorial indexes the condition of total symmetry, one is left with  the expected  number of $2(2j+1)$ independent degrees of freedom. The BW projector is necessarily of the order $\partial ^{2j}$ in the derivatives, and so are the related spin-$j$ wave equations and associated Lagrangians. Differential wave equations of the order higher than two can not be gauged in a consistent manner, and allow  several unphysical aspects, all problematic to deal with,  such as non-locality, acausality, ghosts and etc to enter the theory. In order to avoid appearance of difficulties of this kind we here suggest a strategy of replacing the high order of the BW wave equations by the universal second order.
Such is  achieved in first replacing the BW projector by a projector of zeroth order, $\partial^0$, in the derivatives, which we built up from one of the Casimir invariants of the Lorentz group when exclusively acting on spaces of internal spin degrees of freedom. This projector  allows one to unambiguously and directly  identify anyone of the irreducible sectors of the primordial rank-$2j$ spinor, be it the aforementioned $(j,0)\oplus (0,j)$--, or any other representation space, and  without any reference to the external space-time and the four-momentum.  The dynamics is then introduced by conditioning the projected $(j,0)\oplus (0,j)$  sector  to satisfy the Klein-Gordon equation. The scheme allows for a consistent minimal gauging.
In addition, the Lorentz projector method allows  one to reveal and avoid a shortcoming of the BW framework concerning the  $2(2j+1)$ degrees of freedom, 
which we find to transform reducibly under Lorentz transformations insofar  as they do not exclusively reside in $(j,0)\oplus (0,j)$ but have non-vanishing contributions from other irreducible representation spaces containing the spin-$j$ of interest.
  
\end{flushleft}

{\bf Introduction.} High spin $j\geq 1$ fields, both massive and massless, always have been among the principal topics in field theories. In particle physics they  are needed in the description of the reported hadron resonances whose spins vary from $1/2$ to $17/2$ for baryons, and from zero to six for mesons. In gravity, higher spins can couple to the metric tensor and cause its deformation \cite{Gravity}, and are besides this in demand in the physics of rotating black holes \cite{rot_BH}. Gravitational interactions among high-spin fermions are also under discussion \cite{A}.
Various approaches to high spin fields have been developed over the years (see \cite{HSReview} for a recent review), the Bargmann-Wigner (BW) framework \cite{BW1948} being one of them. In designing high spins as spinors of high ranks, arising from local direct products of multiple Dirac spinors copies, the BW approach is of special interest in gravity in the spinor formalism \cite{Spinorbook}. All the traditional high spin theories have in common that they describe the fields by means of differential wave equations (and associated Lagrangians) which  are of the order $\partial ^{2j}$ in the derivatives.
Differential equations of the order higher than second are difficult to tackle with as they can not be properly gauged, and suffer severe inconsistencies such as non-locality, ghost solutions and acausality in propagation \cite{HSReview}. In view of these difficulties,  finding a strategy of lowering the order of the 
differential high spin equations is timely and of interest. It is the goal of the present work to suggest such a strategy specifically for the BW framework. 
Before, and for the sake of the self-sufficiency of the presentation, we briefly highlight  the method aimed to be remedied.\\

{\bf The Bargmann-Wigner framework.}
The Bargmann-Wigner framework \cite{BW1948} for the description of particles with any spin, $j$ (preceded by a work of de Broglie \cite{deBroigle} for spin-$1$) is based on the general principle for constructing irreducible representation spaces of the Lorentz group (considered in its action
on the internal spin degrees of freedom alone) using its fundamental four-component spinor as a principal building block.
To be specific, as a first step one  constructs an $n$-fold local direct product of fundamental $(1/2,0)\oplus (0, 1/2)$ spinors, with $n=2,3,..$, thus  generating a large Lorentz reducible representation space, which is then decomposed  into a direct sum of irreducible Lorentz group representation spaces containing spins ranging from zero to $n/2$ for even $n$, or, from $1/2$ to $n/2$ for odd $n$'s. Finally, employing a properly designed projector, one pins down one of these irreps, containing the spin of interest.  As we shall see below, the Lorentz group, when limited to  act on the internal spin degrees of freedom alone, provides quite comfortable  tools for catching any irrep in any direct sum of such and without any reference to the external space time and the four-momentum. So far, however, the Bargmann-Wigner framework has pursued a different path, briefly highlighted below.\\ 

\noindent
The direct product, 
\begin{equation}
\Psi^{(n)}_{\alpha_1\alpha_2...\alpha_n}=\otimes _{i=1}^{i=n}\, \left[(1/2,0)\oplus (0,1/2)\right]_i, \quad \alpha_k=1,2,3,4 \quad \forall\,  k,
\label{Gl_prm}
\end{equation} 
of $n$ Dirac spinors,  $\psi_\alpha \simeq (1/2,0)\oplus (0,1/2)$, for $n=2j$,
defines a rank-$n$ four-component Dirac spinor according to, 
\begin{equation}
\Psi^{(n)}_{\alpha_1\alpha_2...\alpha_n}=\left( 
\begin{array}{c}
\psi_1,\alpha_2,\alpha_3...\alpha_{n}\\
\psi_2,\alpha_2,\alpha_3...\alpha_{n}\\
\psi_3,\alpha_2,\alpha_3...\alpha_{n}\\
\psi_4,\alpha_2,\alpha_3...\alpha_{n}
\end{array}
\right).\label{Gl2}
\end{equation}
The latter  decomposes into irreducible Lorentz group representations spaces as,
\begin{eqnarray}
\Psi^{(n)}_{\alpha_1\alpha_2...\alpha_n}&=&\left(0,0\right)\oplus ... \oplus 
\left(\frac{n}{2},0\right)\oplus \left(0,\frac{n}{2}\right)\oplus ...\oplus \left(\frac{1}{2},\frac{n-1}{2} \right) \oplus 
\left(\frac{n-1}{2}, \frac{1}{2} \right)\oplus...\nonumber\\
&\oplus& \left(m\frac{1}{2},m\frac{1}{2}\right)\oplus..., \qquad m \leq \frac{n}{4}.
 \label{Gl1}
\end{eqnarray}
Here, integer spins, i.e. even $n$ values have been considered  for the sake of concreteness and without loss of generality. The case of the half-integer spin (odd $n$ values) will be addressed to the end of the article.
The equation (\ref{Gl1}) shows that there are several irreducible  Lorentz group representation spaces of different dimensionalities  
which contain the spin of interest, such as
the pure-spin $(n/2,0)\oplus (0,n/2)$--, the two-spin $(1/2,(n-1)/2 )\oplus ((n-1)/2,1/2)$--, and the multiple-spin $(n/4,n/4)$ ones. 
Depending on the projector used, one can pin down one of them and formulate a high-spin theory in which the states are described by means of rank-$2j$ spinors
of the type in (\ref{Gl2}). The projector employed by Bargmann and Wigner, here denoted by $ \Pi^{BW}\left(\partial^{2j} \right)$, is given by  the  direct product of the $n$ copies of the Dirac projector, 
$\pi^D\left(\partial\right)$, the covariant parity projector within the Dirac spinor space, and reads,
\begin{eqnarray}
\Pi ^{BW}\left(\partial^{2j}\right)&=&\otimes _{r=1}^{r=n}\left[ \pi^D(\partial)\right]_r, \quad \pi^D_r(\partial) =\left[\frac{i\gamma^\mu\partial _\mu +m}{2m}\right]_r.
\label{Gl3}
\end{eqnarray}
The BW projector imposes the parity and the on-mass-shell  conditions on each one of the spinors in the product, a reason for which
it becomes of  the order $\partial^{n}\equiv \partial^{2j}$  in the derivatives, as indicated in the parenthesis of its notation.
The two conditions  are  very strong  and not quite necessarily, indeed, and as we shall see below, in dropping them, the uncomfortable  high $\partial ^{2j}$ order of the related  differential wave equations becomes avoidable.  Finally, in order to end up with $2(2j+1)$ degrees of freedom, as required by
$(j,0)\oplus (0,j)$, only totally symmetric spinor indexes need to be considered. 
In this fashion, spin-$j=n/2$ is described within the BW framework by means of a totally symmetric rank-$2j$ four-component spinor by means of 
differential equations of the order $\partial^{2j}$ in the derivatives. As remarked above,
such equations present serious difficulties in so far as they allow for unphysical solutions which need a special effort to be excluded.
Within this context, a re-formulation  of the BW  framework in terms  of differential equations of the regular second order, $\partial^{2}$, is timely and of interest. Our case is that the  Lorentz group, when acting exclusively on the internal spin degrees of freedom,
 does indeed provide the adequate  tools for the realization of such a re-formulation.\\

{\bf Lorentz group transformations of the spin degrees of freedom.}
The Lorentz group transforming the internal spin degrees of freedom, henceforth termed to as ``internal'' Lorentz group, and denoted by, ${\mathcal L}$,  is a subgroup of the complete Lorentz group, the latter acting besides on the spin- also  on the external space time. 
The ${\mathcal L}$ generators, denoted by $S_{\mu\nu}$, are quadratic $d\times d$ constant matrices, where $d$ fixes the finite dimensionality of the internal representation space, and encodes the spin value. For the special case of a pure spin, dimensionality and spin are related as
 $d=2(2j+1)$, while for representations of multiple spins, relations like  $d=\sum_i(2j_i+1)$, or, $ d=2\sum_i (2j_i+1)$, can hold valid. 
 Now, the direct product of ${\mathcal L}$ and  ${\mathcal T}_4$, the group of translations in the external space-time,
  whose generators, $i\partial _\mu$,  represent the quantum mechanical operators, $P_\mu$,   
of the components of the relativistic four-momentum, i.e. $P_\mu=i\partial_\mu$,  is generated by the following sub-set of the Poincar\'e algebra:
\begin{eqnarray}
{\mathcal L}:\quad \left[S_{\mu\nu},S_{\rho\sigma}\right] &=&i(g_{\mu\rho}S_{\nu \sigma}-g_{\nu\rho}S_{\mu\sigma}+g_{\mu\sigma}S_{\rho\nu}-g_{\nu\sigma}S_{\rho\mu}),\label{Lrntz_algbr}\\
{\mathcal T}:\qquad \left[ P_\mu,P_\nu \right]&=&0,\label{trnsl_algbr}\\
 \left[S_{\mu\nu},P_\lambda \right] &=&0.
\label{intrtwn}
\end{eqnarray}
In the algebra of the full Poincar\'e group the commutators in (\ref{intrtwn}) are non-vanishing because there,
 the Lorentz group generators in the internal space, the  constant matrices $S_{\mu\nu}$, are supplemented
by the angular momentum and boost operators, $L_{\mu\nu}=-i\left(x_\mu \partial ^\nu  -x_\nu \partial^\mu \right)$, 
which transform the external space time, and which do not commute with the operators of translations.
Nonetheless, the ${\mathcal T}_4\times {\mathcal L}$ group algebra has same Casimir invariants as the full Poincar\'e group algebra \cite{KimNoz}. These 
invariants refer to the operators of the  squared four-momentum, $P^2$, and  the square of the  Pauli-Lubanski vector, 
${\mathcal W}^\mu$, defined as,
\begin{eqnarray}
\left[W^\mu\right]_{AB}&=&\frac{1}{2}\epsilon_{\lambda \rho\sigma \mu} \left[ S^{\rho\sigma}\right]_{AB}P^\mu,
\label{PL}\\
A,B, C, D, ...&=&1, 2,..,d.
\label{dimspace}
\end{eqnarray}
Here, the  capital Latin letters label the generic internal finite $d$-dimensional representation space of interest (it can be either reducible, or, irreducible).
On the other side, the algebra of the internal Lorentz group, given in (\ref{Lrntz_algbr}),  has by itself two Casimir invariants  \cite{KimNoz},
here denoted in their turn by, $F$ and $G$, and  defined as,
\begin{eqnarray}
F_{AB}&=& \frac{1}{2}\left[S^{\mu\nu}\right]_{AD}\left[S_{\mu\nu}\right]_{DB}, \label{Cas1}\\ 
G_{AB}&=&\frac{1}{2}\epsilon_{\mu\nu\alpha\beta}\left[S^{\mu\nu}\right]_{AC}\left[S_{\alpha\beta}\right]_{CB}.
\label{as2}
\end{eqnarray}
Due to the constancy of the quadratic $d\times d$ dimensional matrices $S_{\mu\nu}$, the  $F$, and $G$ operators commute with both $P^2$, and ${\mathcal W}^2$, without by themselves  being Poincar\'e  invariants. 
This is a remarkable property which we will put at work in what follows.
These two operators, have the property of unambiguously identifying  any \underline{irreducible}
finite dimensional ${\mathcal L}$ group representation space, here generically denoted by, $(j_1,j_2)\oplus (j_2,j_1)$,  through their eigenvalues according to,
\begin{eqnarray}
F \, (j_1,j_2)\oplus (j_2,j_1)&=&c_{(j_1,j_2)} \left(j_1,j_2 \right)\oplus (j_2,j_1),
\label{F_Cas}\\
c_{(j_1,j_2)}&=&\frac{1}{2}\left(K(K+2)+M^2\right), \quad K=j_1+j_2, \quad M=|j_1-j_2|,
\label{F_csts}\\
G\, \, (j_1,j_2)\oplus (j_2,j_1)&=&r_{(j_1,j_2)}\left(j_1,j_2 \right)\oplus (j_2,j_1),
\label{G_Cas}\\
r_{(j_1,j_2)}&=&iKM.
\label{G_csts}
\end{eqnarray}
The idea of the present work, executed in the next section, is to employ the Casimir invariant $F$ in the construction of a momentum independent (static) projector on the irreducible sectors of the $n=2j$ rank spinor $\Psi^{(n)}_{\alpha_1\alpha_2...\alpha_n}$ in (\ref{Gl1}) and to explore the consequences. The article closes with brief conclusions.\\

\noindent
{\bf Static projector on the $(n/2,0)\oplus (0,n/2)$ irreducible sector in the Bargmann-Wigner rank-$(2j)$ spinor.} We here are specifically interested in projectors on the irreducible Lorentz representations  appearing in the rhs of the equation (\ref{Gl1}) which contain  spin $j=n/2$.
The first  projector we wish to consider, here denoted by ${\mathcal P}^{(n/2,0)}$, is the one that identifies the $(n/2,0)\oplus (0,n/2)$ irrep.
We construct it from $F$ in (\ref{F_Cas}) as
\begin{equation}
\mathcal{P}_F^{(n/2,0)}= \Pi_{k l} \times \left( \dfrac{F-c_{(j_k,j_l)}}{c_{(n/2,0)}-c_{(j_k,j_l)}} \right), \label{General_L_Proj}
\end{equation}
where, $\Pi_{kl} \times$ denotes the ordinary product of the  operators in parenthesis, the pairs of indexes $(k,l)$ run over all the $(j_k,j_l)$ labels characterizing  the  irreducible
representation spaces in the rhs of (\ref{Gl1}), and the constants  $ c_{(j_k,j_l)}$ 
are those defined in (\ref{F_csts}). The equation (\ref{General_L_Proj})
shows that the operator $\mathcal{P}_F^{(n/2,0)}$ has the property to nullify any irreducible representation space for which
$(j_k,j_l)\not= (n/2,0)$. Instead, for $(j_k,j_l) = (n/2,0)$, it acts as the identity operator, meaning that $\mathcal{P}_F^{(n/2,0)}$
is a projector on $(n/2,0)\oplus (0,n/2)$.
It is obvious that $\mathcal{P}_F^{(n/2,0)}$ is of zeroth order in the momenta (the derivatives).
Such is possible due to the product character, ${\mathcal T}_4\times {\mathcal L}$ of the symmetry group.
Indeed, the internal Lorentz group ${\mathcal L}$, in factorizing from the group of translations, and in exclusively acting on the spin space, 
does not need external transformations for the identification of its irreducible  degrees of freedom.
Furthermore, it is sufficient to require the $(n/2,0)\oplus (0,n/2)$ states to be on their mass-shell, and drop
the requirement of good parities and on-mass-shell-ness of the ``constituent'' Dirac spinors. The reward will be a second order differential 
realization within  the Bargmann-Wigner basis. 
In the following we re-formulate the BW framework in terms of ${\mathcal P}_F^{(n/2,0)}$, and explore consequences. \\

\noindent
{\bf Second order differential realization of the Bargmann-Wigner framework.}
The  Lorentz group generators, $\left[ S^{(1)} \right]_{\mu \nu}$,  in  the Dirac spinor space, $(1/2,0)\oplus(0,1/2)$, are textbook knowledge and read,
\begin{equation}
\left[ S^{(1)} \right]_{\mu \nu}=\dfrac{i}{4}[\gamma_\mu, \gamma_\nu]=\dfrac{1}{2} \sigma_{\mu \nu}, \quad \mu, \nu=0,1,2,3,
\label{Spinor_generators}
\end{equation}
where $\gamma_{\mu}$, and $\gamma_\nu$ stand for the standard Dirac matrices. Then,  the internal Lorentz group generators, $\left[ S^{(n)} \right]_{\mu \nu}$,  
in the reducible Bargmann-Wigner $\Psi^{(n)}_{\alpha_1\alpha_2...\alpha_n}=\otimes _{i=1}^{i=n} \left[(1/2,0)\oplus (0,1/2)\right]_i$ basis in (\ref{Gl1}),  are calculated following the standard prescription regarding generator construction in product spaces which is,
\begin{equation}
\left[ S^{(n)} \right]_{\mu \nu}=\left[ S^{(1)} \right]_{\mu \nu} \otimes \left[ \Pi_{i=1}^{i=n-1} \otimes \mathbf{1}\right]
 + \mathbf{1} \otimes \left[ S^{(1)} \right]_{\mu \nu}\left[  \otimes \Pi_{i=1}^{i=n-2} \otimes \mathbf{1}\right]+ \cdots + 
\left[ \Pi_{i=1}^{i=n-1} \otimes \mathbf{1}\right] \otimes \left[ S^{(1)} \right]_{\mu \nu},
\label{Generator_direct_product}
\end{equation}
with $\left[ S^{(1)} \right]_{\mu \nu}$ in (\ref{Spinor_generators}), and  $\mathbf{1}$ standing for the $4 \times 4$ unit matrix in the Dirac spinor space.  In substituting  (\ref{Spinor_generators}) into (\ref{Generator_direct_product}), and then (\ref{Generator_direct_product}) into (\ref{General_L_Proj}),
the $P_F^{(n/2,0)}$ projector of interest is explicitly constructed. Once again, it is seen that this  projector is a static one and does not provide a wave equation. In order to introduce the dynamics, we impose on the states spanning the $(n/2,0)\oplus (0,n/2)$ representation space the mass-shell condition via the Klein-Gordon equation, and find the following master equation,
\begin{equation}
\left( \partial_\mu \partial^\mu + m^2 \right) \left[ \mathcal{P}^{(n/2,0)}_{F} \right]_{\alpha_1 \alpha_2\dots \alpha_{n}}^{~~~~~~~~~~~~~~~ \beta_1 \beta_2 \dots \beta_{n}} \Psi^{(n)}_{\beta_1 \beta_2 \dots \beta_{n}}=0, \quad n=2j. \label{second_order_equation_spinorialtensor}
\end{equation}
The latter equation can be coupled to the electromagnetic field in the regular way and without creating problems.  
In this manner, we furnished a second order differential wave equation for particles of spin-$j=n/2$ described in terms of  a Dirac spinor of rank $2j$. In the following, we work out for illustrative purposes  
the case of spin-$1$ residing in the $(1,0)\oplus (0,1)$ irreducible sector of Dirac spinor, $\Psi^{(2)}$, of second rank.\\

\noindent
{\bf The wave equation of spin-$1$ in $\Psi^{(2)}$. An illustrative example.}
As an illustrative example for our suggested method we consider the simplest case of an integer spin, namely,  spin-$1$ as part of the Dirac spinor of second rank,
$\Psi^{(2)}_{\alpha_1\alpha_2}=\psi_{\alpha_1}\psi_{\alpha_2}\simeq \left[(1/2,0)\oplus (0,1/2)\right]\otimes \left (1/2,0)\oplus (0,1/2)\right]$. 
The latter tensor is reducible under Lorentz transformations according to, 
\begin{equation}
\Psi^{(2)}_{\alpha_1\alpha_2}\simeq [(1/2,0)\oplus(0,1/2)]\otimes [(1/2,0)\oplus(0,1/2)]=(1,0)\oplus(0,1)\oplus 2(0,0) \oplus 2(1/2,1/2),
\label{spin1_1}
\end{equation}
where the integer numbers in front of the irreps stand for their multiplicities in the direct sum.
The rhs in (\ref{spin1_1}) contains three different irreducible Lorentz representation spaces, whose eigenvalues with respect to the Casimir invariant $F$ in (\ref{F_Cas}) are calculated from the expressions given in (\ref{F_csts}) as,
\begin{eqnarray}
c_{(0,0)} &=&0 \quad \quad \text{for} \quad (0,0), \\
c_{(1,0)} &=&2 \quad \quad \text{for} \quad (1,0)\oplus(0,1), \\
c_{(1/2,1/2)} &=&\frac{3}{2} \quad \quad \text{for} \quad (1/2,1/2).
\end{eqnarray}
Therefore, according to (\ref{General_L_Proj}), the projector ${\mathcal P}_F^{(1,0)}$ on $(1,0)\oplus (0,1)$  emerges as,
\begin{equation}
\mathcal{P}_F^{(1,0)}=\dfrac{1}{2}\left( 2F^2-3F \right). \label{Lor_proj_form1}
\end{equation}
Next we evaluate  the generators within $\Psi^{(2)}_{\alpha \beta}$  by the aid of (\ref{Generator_direct_product}) setting  $n=2$, 
substitute them  in (\ref{Cas1}), and then insert the result for $F$ in   (\ref{General_L_Proj}).
In so doing we calculate the following explicit expression for the $F$- Casimir invariant of the internal Lorentz algebra: 
\begin{eqnarray}
\Psi_{\alpha\beta}^{(2)}:\quad F&=&\dfrac{1}{16} \left( \mathbf{1} \otimes \sigma^{\mu \nu} + \sigma^{\mu \nu} \otimes \mathbf{1} \right)
\left( \mathbf{1} \otimes \sigma_{\mu \nu} + \sigma_{\mu \nu} \otimes \mathbf{1} \right).
\label{spin1_2}
\end{eqnarray}
The tensor form of  $F$ is now given by, 
\begin{equation}
F^{\alpha_1 \alpha_2}_{~~~~~~~~\beta_1 \beta_2} =\dfrac{1}{8}\left[ 12\; \delta^{\alpha_1}_{~~~~\beta_1} \delta^{\alpha_2}_{~~~~\beta_2} + \left( \sigma_{\mu \nu} \right)^{\alpha_1}_{~~~~\beta_1} \left( \sigma^{\mu \nu} \right)^{\alpha_2}_{~~~~\beta_2} \right]. \label{F_spinor_components}
\end{equation}
Substitution of (\ref{F_spinor_components}) into (\ref{Lor_proj_form1}) amounts to the following explicit expression for the
searched spin-$1$ Lorentz projector:
\begin{eqnarray}
\left[ \mathcal{P}_F^{(1,0)} \right]^{\alpha_1 \alpha_2}_{~~~~~~~~\beta_1 \beta_2}&=&\dfrac{1}{32}
\left( \sigma_{\mu \nu} \right)^{\alpha_1}_{~~~~\beta'_1}\left( \sigma_{\mu \nu} \right)^{\alpha_2}_{~~~~\beta'_2} \left[ 12\; \delta^{\beta'_1}_{~~~~\beta_1} \delta^{\beta'_2}_{~~~~\beta_2} + \left( \sigma_{\mu \nu} \right)^{\beta'_1}_{~~~~\beta_1} \left( \sigma^{\mu \nu} \right)^{\beta'_2}_{~~~~\beta_2} \right] \label{Lor_proj_form_components}  \\ \nonumber  
&=& \dfrac{1}{4}
\left( \sigma_{\mu \nu} \right)^{\alpha_1}_{~~~~\beta'_1}\left( \sigma_{\mu \nu} \right)^{\alpha_2}_{~~~~\beta'_2}
 F^{\beta'_1 \beta'_2}_{~~~~~~~~\beta_1 \beta_2}.
\end{eqnarray}
Therefore, according to (\ref{second_order_equation_spinorialtensor}), the second order differential  wave equation for a particle of spin-$1$ 
described in terms of a second rank Dirac spinor takes the following form:
\begin{equation}
\left( \partial_\mu \partial^\mu + m^2 \right)
\left[ \mathcal{P}_F^{(1,0)} \right]_{\alpha_1 \alpha_2}^{~~~~~~~~\beta_1 \beta_2}
 \Psi^{(2)}_{\beta_1 \beta_2}=0. \label{Wave_equation_spin1}
\end{equation}
Along this line, any arbitrary spin-$j$ can be described by means of a rank-$2j$ Dirac spinor satisfying a second order differential equation.
Below the solutions to the equation (\ref{Wave_equation_spin1}), when considered in the momentum space upon the replacement, $i\partial_\mu \to P_\mu$,  are obtained, and compared to the corresponding  solutions appearing in the  Bargmann-Wigner framework.\\

\noindent
{\bf Comparison of the spin-$1$  Lorentz--,  and  Bargmann-Wigner projections  on $\Psi^{(2)}$.}
A set of linearly independent solutions in  momentum space for the spin-$1$ states satisfying  the equation in (\ref{Wave_equation_spin1}) 
can be constructed upon applying the Lorentz projector $\mathcal{P}_F^{(1,0)}$ in (\ref{Lor_proj_form_components}) to the sixteen dimensional
rank-2 Dirac spinors, $\left[ u_\pm(\mathbf{p},\lambda) \right]^\alpha \left[ u_\pm(\mathbf{p},\lambda) \right]^\beta$, as composed by the momentum space
Dirac spinors of positive ($+$),  and negative, ($-$), parities,
\begin{eqnarray}
u_{+}(\mathbf{p},1/2)\equiv  u(\mathbf{p},1/2)&=&\dfrac{1}{\sqrt{2m(m+p_0)}}\begin{pmatrix}
m+p_0 \\ 0 \\ p_3\\ p_1+ip_2
\end{pmatrix}, \nonumber\\\ 
u+(\mathbf{p},-1/2)\equiv u(\mathbf{p},-1/2)&=&\dfrac{1}{\sqrt{2m(m+p_0)}}\begin{pmatrix}
0\\ m+p_0\\ p_1-ip_2 \\ -p_3
\end{pmatrix}, \nonumber\\
u_{-}(\mathbf{p},1/2)\equiv v(\mathbf{p},1/2)&=&\dfrac{1}{\sqrt{2m(m+p_0)}} \begin{pmatrix}
p_3\\p_1+ip_2\\m+p_0\\0
\end{pmatrix},\nonumber\\
u_{-}(\mathbf{p},-1/2)\equiv v(\mathbf{p},-1/2)&=&\dfrac{1}{\sqrt{2m(m+p_0)}} \begin{pmatrix}
p_1-ip_2\\-p_3\\0\\m+p_0
\end{pmatrix}.
\end{eqnarray}
The sixteen dimensional rank-2 Dirac spinors  span the  reducible Lorentz representation space $[(1/2,0) \oplus (0,1/2)] \otimes [(1/2,0) \oplus (0,1/2)]$. Executing the proper calculation, we find precisely  six linearly independent combinations, as it should be, and  list them in the Table \ref{Comparative_table_basis} together with the corresponding Bargmann-Wigner spin-$1$ states, as a comparison.

\begin{table}[h]
\begin{tabular}{|c | c|}
\hline
Normalized spin-$1$  states (this work) & Normalized spin-$1$ Bargmann-Wigner states \\ [0.5ex] \hline

$\begin{array}{l c l}
\left[ w_+(\mathbf{p},1) \right]^{\alpha \beta}&=&\frac{1}{\sqrt{2}} \big( \left[ u_+(\mathbf{p},1/2) \right]^\alpha \left[ u_+(\mathbf{p},1/2) \right]^\beta \\
& & + \left[ u_-(\mathbf{p},1/2) \right]^\alpha \left[ u_-(\mathbf{p},1/2) \right]^\beta \big)
\end{array}$ & 
$\left[ w_+^{(BW)}(\mathbf{p},1) \right]^{\alpha \beta}= \left[ u_+(\mathbf{p},1/2) \right]^\alpha \left[ u_+(\mathbf{p},1/2) \right]^\beta$ \\ [0.5cm] \hline

$\begin{array}{l c l}
\left[ w_+(\mathbf{p},0) \right]^{\alpha \beta}&=&\frac{1}{2} \big( \left[ u_+(\mathbf{p},1/2) \right]^\alpha \left[ u_+(\mathbf{p},-1/2) \right]^\beta \\
& & + \left[ u_+(\mathbf{p},-1/2) \right]^\alpha \left[ u_+(\mathbf{p},1/2) \right]^\beta \\
& & + \left[ u_-(\mathbf{p},1/2) \right]^\alpha \left[ u_-(\mathbf{p},-1/2) \right]^\beta \\
& & + \left[ u_-(\mathbf{p},-1/2) \right]^\alpha \left[ u_-(\mathbf{p},1/2) \right]^\beta \big)
\end{array}$ & 
$\begin{array}{l c l}
\left[ w_+^{(BW)}(\mathbf{p},0) \right]^{\alpha \beta}&=& \left[ u_+(\mathbf{p},1/2) \right]^\alpha \left[ u_+(\mathbf{p},-1/2) \right]^\beta \\
& & + \left[ u_+(\mathbf{p},-1/2) \right]^\alpha \left[ u_+(\mathbf{p},1/2) \right]^\beta 
\end{array}$\\ [1cm] \hline

$\begin{array}{l c l}
\left[ w_+(\mathbf{p},-1) \right]^{\alpha \beta}&=&\frac{1}{\sqrt{2}} \big( \left[ u_+(\mathbf{p},-1/2) \right]^\alpha \left[ u_+(\mathbf{p},-1/2) \right]^\beta \\
& & + \left[ u_-(\mathbf{p},-1/2) \right]^\alpha \left[ u_-(\mathbf{p},-1/2) \right]^\beta \big)
\end{array}$ & 
$\left[ w_+^{(BW)}(\mathbf{p},-1) \right]^{\alpha \beta}= \left[ u_+(\mathbf{p},-1/2) \right]^\alpha \left[ u_+(\mathbf{p},-1/2) \right]^\beta$ \\ [0.5cm] \hline

$\begin{array}{l c l}
\left[ w_-(\mathbf{p},1) \right]^{\alpha \beta}&=&\frac{1}{\sqrt{2}} \big( \left[ u_+(\mathbf{p},1/2) \right]^\alpha \left[ u_-(\mathbf{p},1/2) \right]^\beta \\
& & + \left[ u_-(\mathbf{p},1/2) \right]^\alpha \left[ u_+(\mathbf{p},1/2) \right]^\beta \big)
\end{array}$ & 
$\left[ {\widetilde w}_+^{(BW)}(\mathbf{p},1) \right]^{\alpha \beta}= \left[ u_-(\mathbf{p},1/2) \right]^\alpha \left[ u_-(\mathbf{p},1/2) \right]^\beta$ \\ [0.5cm] \hline

$\begin{array}{l c l}
\left[ w_-(\mathbf{p},0) \right]^{\alpha \beta}&=&\frac{1}{2} \big( \left[ u_+(\mathbf{p},1/2) \right]^\alpha \left[ u_-(\mathbf{p},-1/2) \right]^\beta \\
& & + \left[ u_-(\mathbf{p},-1/2) \right]^\alpha \left[ u_+(\mathbf{p},1/2) \right]^\beta \\
& & + \left[ u_+(\mathbf{p},1/2) \right]^\alpha \left[ u_-(\mathbf{p},-1/2) \right]^\beta \\
& & + \left[ u_-(\mathbf{p},-1/2) \right]^\alpha \left[ u_+(\mathbf{p},1/2) \right]^\beta \big)
\end{array}$ & 
$\begin{array}{l c l}
\left[ {\widetilde w}_+^{(BW)}(\mathbf{p},0) \right]^{\alpha \beta}&=& \left[ u_-(\mathbf{p},1/2) \right]^\alpha \left[ u_-(\mathbf{p},-1/2) \right]^\beta \\
& & + \left[ u_-(\mathbf{p},-1/2) \right]^\alpha \left[ u_-(\mathbf{p},1/2) \right]^\beta 
\end{array}$\\ [1cm] \hline

$\begin{array}{l c l}
\left[ w_-(\mathbf{p},-1) \right]^{\alpha \beta}&=&\frac{1}{\sqrt{2}} \big( \left[ u_+(\mathbf{p},-1/2) \right]^\alpha \left[ u_-(\mathbf{p},-1/2) \right]^\beta \\
& & + \left[ u_-(\mathbf{p},-1/2) \right]^\alpha \left[ u_+(\mathbf{p},-1/2) \right]^\beta \big)
\end{array}$ & 
$\left[ {\widetilde w}_+^{(BW)}(\mathbf{p},-1) \right]^{\alpha \beta}= \left[ u_-(\mathbf{p},-1/2) \right]^\alpha \left[ u_-(\mathbf{p},-1/2) \right]^\beta$ \\ [0.5cm] \hline
\end{tabular}
\caption{Comparison  between the spin-$1$ states obtained in this work (left column) with the spin-$1$ states predicted by the Bargmann-Wigner equations (right column).}
\label{Comparative_table_basis}
\end{table}

\noindent
In order to make the comparison between the two schemes manifest, we write down in the subsequent two equations the explicit rank-$2$ Dirac spinors following from the present work, $\left[w_\pm ({\mathbf p},\lambda) \right]^{\alpha\beta}$, on the one side, and from the Bargmann-Wigner approach, 
$\left[w_+ ^{(BW)}({\mathbf p},\lambda) \right]^{\alpha\beta}$-$\left[{\widetilde w}_+ ^{(BW)}({\mathbf p},\lambda) \right]^{\alpha\beta}$, on the other, focusing on the particular case of $\left[w_{+}({\mathbf p},1)\right]^{\alpha\beta}$, and $\left[w_{+}^{(BW)}({\mathbf p},1)\right]^{\alpha\beta}$. In so doing, we find,

\begin{eqnarray}
\left[w_{+}({\mathbf p},1)\right]^{\alpha\beta}&=&\sqrt{2}\left( 
\begin{array}{cccc}
\frac{(m+p_0)^2 +p_3^2}{4m(m+p_0)}& \frac{(p_1+ip_2)p_3}{4m(m+p_0)}& \frac{p_3}{2m}&\frac{p_1+ip_2}{4m}\\
 \frac{(p_1+ip_2)p_3}{4m(m+p_0)}&\frac{(p_1+ip_2)^2}{4m(m+p_0)}&\frac{p_1+ip_2}{4m}&0\\
\frac{p_3}{2m}&\frac{p_1+ip_2}{4m}&\frac{(m+p_0)^2 +p_3^2}{4m(m+p_0)}& \frac{(p_1+ip_2)p_3}{4m(m+p_0)}\\
\frac{p_1+ip_2}{4m}&0&\frac{(p_1+ip_2)p_3}{4m(m+p_0)}&\frac{(p_1+ip_2)^2}{4m(m+p_0)}
\end{array}
\right),
\label{we}
\end{eqnarray}
and

\begin{eqnarray}
\left[w_{+}{}^{(BW)}({\mathbf p},1)\right]^{\alpha\beta}&=&\frac{1}{2m(m+p_0)}\left(
\begin{array}{cccc}
(m+p_0)^2& 0&p_3(m+p_0)&(m+p_0)(p_1+ip_2)\\
0&0&0&0\\
p_3(m+p_0)&0&p_0^2&p_3(p_1+ip_2)\\
(p_1+ip_2)(m+p_0)&0&p_3(p_1+ip_2)&(p_1+ip_2)
\end{array}\right).\nonumber\\
\label{they}
\end{eqnarray}

Inspection of these equations shows that the tensor calculated in the equation (\ref{we}) of the present work  has the following six independent degrees of freedom:
\begin{eqnarray}
w^{11}=w^{33}, &\quad&  w^{12}=w^{21}=w^{34}=w^{43}, \quad w^{13}=w^{31}, \nonumber\\
w^{22}=w^{44}, &\quad& w^{14}=w^{41}=w^{23}=w^{32}, \quad w^{24}=w^{42}, 
\label{dof_we}
\end{eqnarray}
and so does the Bargmann-Wigner solution in (\ref{they}), though theirs are distinct from ours. The six independent spin-$1$ degrees of freedom  
following from the Bargmann-Wigner framework are,
\begin{eqnarray}
w^{(BW)}{}^{11}, &\quad& w^{(BW)}{}^{13}=w^{(BW)}{}^{31}, \quad w^{(BW)}{}^{14}=w^{(BW)}{}^{41}, \nonumber\\
w^{(BW)}{}^{33}, &\quad& w^{(BW)}{}^{34}=w^{(BW)}{}^{43}, \quad w^{(BW)}{}^{44},
\label{dof_they}
\end{eqnarray}
while the rest of the components is identically vanishing.
In the last two equations we dropped  subscripts  and arguments in the $\left[ w_{+}({\mathbf p },1)\right]^{\alpha\beta}$ spinor components under consideration for the sake of keeping the notations possibly  more transparent.
In a similar way, the remaining $\Psi^{(2)}_{\alpha\beta}$ momentum space components can be analyzed.
However, though the spinors obtained in the two approaches have six degrees of freedom each, they are not identical.
In the following we make the case that the Bargmann-Wigner spin-$1$ spinors transform reducibly under Lorentz transformations, while our spinors transform irreducibly, as it should be if the spin under consideration were to reside entirely in $(1,0)\oplus (0,1)$.
{}For this purpose we construct the Lorentz projector, ${\mathcal P}_F^{(1/2,1/2)}$, on  the $(1/2,1/2)$ irreducible sector in the rhs in (\ref{Gl1})
and let it act on the Bargmann-Wigner spin-$1$ spinors.\\

\noindent
{\bf The Lorentz projector on the irreducible $(1/2,1/2)$ sector in $\Psi^{(2)}$.}
Without entering into technical details, we limit ourselves to report  that the projector of our interest is obtained along the line of the 
reasoning presented above and its general form is found as,
\begin{equation}
\mathcal{P}_F^{(1/2,1/2)}=\dfrac{4}{3} ( 2F-F^2 ).
\label{4vprjct}
\end{equation}
In spinor index notation we calculated it as, 
\begin{equation}
\left[ \mathcal{P}_F^{(1/2,1/2)} \right]^{\alpha_1 \alpha_2}_{~~~~~~~ \beta_1 \beta_2} = \delta^{\alpha_1}_{~~~~ \beta_1} \delta^{\alpha_2}_{~~~~ \beta_2}
-\dfrac{1}{12} \left( \sigma_{\mu_1 \nu_1}  \right)^{\alpha_1}_{~~~~ \eta} \left( \sigma_{\mu_2 \nu_2}  \right)^{\eta}_{~~~~ \beta_1} 
\left( \sigma^{\mu_1 \nu_1}  \right)^{\alpha_1}_{~~~~ \rho} \left( \sigma^{\mu_2 \nu_2}  \right)^{\rho}_{~~~~ \beta_2}-
\dfrac{2}{12} \left( \sigma_{\mu \nu} \right)^{\alpha_1}_{~~~~ \beta_1} \left( \sigma^{\mu \nu} \right)^{\alpha_2}_{~~~~ \beta_2}.
\label{4v_spnindx}
\end{equation}

When applied to anyone of the  $\left[ w_+^{(BW)}(\mathbf{p},\lambda) \right]^{\alpha\beta}{}$-
$\left[{\widetilde w}_+ ^{(BW)}({\mathbf p},\lambda) \right]^{\alpha\beta}$ spinors, non-vanishing  projections on
$(1/2,1/2)$ states  (not shown here) are obtained. This means that the spin-$1$ states following from the  Bargmann-Wigner framework
 are  linear combinations of spin-$1$ states residing in  the two distinct Lorentz irreducible representation spaces, $(1,0)\oplus(0,1)$, and 
$(1/2,1/2)$. In other words, differently from our spin-$1$ degrees of freedom, those obtained from the BW approach do not transform irreducibly under Lorentz transformations,  a shortcoming which could prejudice the quantization procedure.\\

\noindent
{\bf Conclusions.} In the present work we suggested an approach to the description of  any spin-$j$  by means of second order differential equations and within the Bargmann-Wigner basis of rank-$2j$ Dirac spinors. Though we here considered integer spin as illustrative of our method, 
the scheme is universal and applies equally well to massive parity mixed (chiral) fermions.
It furthermore applies to any reducible basis embedding the spin $(j,0)\oplus (0,j)$  of interest and generated by direct products of any arbitrary finite dimensional irreducible Lorenz representations spaces. Examples for such bases are provided by  
$n$-fold local direct products of be it four-vectors, $A_\mu\simeq (1/2,1/2)$, or  second rank Lorentz tensors, $B_{\mu\nu}\simeq (1,0)\oplus (0,1)$ for bosons, as well as by  bases obtained from the previous ones  upon taking their direct products with a Dirac spinor, for fermions \cite{AGK}.
The virtue of the method, besides skipping  the problems of the inconsistencies of the high order differential equations, innate to the traditional methods for high-spin description is, that it presents itself computer friendly with respect to symbolic software such as Mathematica and FeynCalc.

\end{document}